\documentclass{article}

\usepackage{arxiv}

\usepackage[utf8]{inputenc} 
\usepackage[T1]{fontenc}    
\usepackage{hyperref}       
\usepackage{url}            
\usepackage{booktabs}       
\usepackage{amsfonts}       
\usepackage{nicefrac}       
\usepackage{microtype}      
\usepackage{lipsum}

\usepackage{multicol}
\usepackage{graphicx}
\usepackage{caption}
\usepackage{subcaption}
\usepackage{multirow}
\usepackage{enumitem}

\title{A Robust Blockchain Readiness Index Model}

\author{
  Elias Iosif\\
  Department of Digital Innovation, School of Business\\
  Institute For the Future (IFF)\\
  University of Nicosia\\
  Cyprus\\
  \texttt{iosif.e@unic.ac.cy}\\
   \And
  Klitos Christodoulou\\
  Department of Digital Innovation, School of Business\\
  Institute For the Future (IFF)\\
  University of Nicosia\\
  Cyprus\\
  \texttt{christodoulou.kl@unic.ac.cy}\\
  \And
  Andreas Vlachos\\
  Department of Digital Innovation, School of Business\\
  University of Nicosia\\
  Cyprus\\
  \texttt{vlachos.a@unic.ac.cy}\\
}

\begin{document}
\maketitle

\newcommand{\eg}{e.g., } 
\newcommand{\ie}{i.e., } 
\newcommand{\cf}[1]{{ cf. #1}}
\newcommand{\etal}{{et al.}\ }
\newcommand{\quotes}[1]{``#1''}

\begin{abstract}
As\footnote{
The final authenticated version is available online at
\url{https://doi.org/10.1007/978-3-030-95947-0_7}
}
the blockchain ecosystem gets more mature many businesses, investors, and entrepreneurs are seeking opportunities on working with blockchain systems and cryptocurrencies. A critical challenge for these actors is to identify the most suitable environment to start or evolve their businesses. In general, the question is to identify which countries are offering the most suitable conditions to host their blockchain-based activities and implement their innovative projects. The Blockchain Readiness Index (BRI) provides a numerical metric (referred to as the blockchain \textit{readiness score}) in measuring the maturity/readiness levels of a country in adopting blockchain and cryptocurrencies. In doing so, BRI leverages on techniques from information retrieval to algorithmically derive an index ranking for a set of countries. The index considers a range of indicators organized under five pillars: Government Regulation, Research, Technology, Industry, and User Engagement. In this paper, we further extent BRI with the capability of deriving the index -- at the country level -- even in the presence of missing information for the indicators. In doing so, we are proposing two weighting schemes namely, linear and sigmoid weighting for refining the initial estimates for the indicator values. A classification framework was employed to evaluate the effectiveness of the developed techniques which yielded to a significant classification accuracy.
\end{abstract}


\section{Introduction}
\label{s:intro}

Blockchain technology has contributed several interesting properties, not only in deriving peer-to-peer software architectures but also in enabling alternative business and transaction models. The blockchain ecosystem evolved from the idea of establishing decentralized ``trust'' \cite{shin2019blockchain}. A decentralized model where multiple distrusting actors, with different motives, are competing with each other in a decentralized, transparent environment. Since the inception of the decentralized trust idea as proposed in \cite{nakamoto2008bitcoin}, blockchain technology and cryptocurrencies have seen a massive adoption in the cyberspace \cite{Makridakis_2019}. 
Despite the adoption and the evolving landscape rooted on social and business interactions among many actors, blockchain technology has been criticized by many nations, governments, law-makers, and financial institutions. However, the community and the dynamic ecosystem reinforced by many social relationships have driven the technology in a long-term evolving state. At the beginning of this disruptive technology this state was deeply rooted in the development of the technology and its advancements, such as, scalability, transaction speeds and interoperability \cite{zheng2017overview}. However, as the adoption of the technology is fueling a variety of applications across many industries \cite{casino2019systematic,christodoulou2018decentralized}, along with many economic, social, and business activities being disrupted, several new and different kinds of challenges are coming into play. 

These challenges underscore the focus to the operational and regulatory dimensions that follow from the widespread of the technology and market adoption. Under this state of affairs, many nations are showing interest in adopting the technology but still they respond differently, and face many challenges in harnessing the potential of this technological innovation \cite{shin2020socio}. On the negative side of the spectrum, many countries have issued a direct ban of the technology and cryptocurrencies, others have taken a more dichotomous approach of the Blockchain technology and cryptocurrencies with limited forms of regulation, and on the positive side of the spectrum countries have taken a more proactive strategy seeking opportunities in becoming ``enabling environments'' for embracing the technology. The following observations motivated this study:

\begin{itemize}
    \item \textit{Governmental authorities often fail to capture the set of variables\footnote{In this work we refer to such variables or factors as \textit{indicators}.} that measure a country's capacity to adopt blockchain and cryptocurrencies as a technological advancement.}
    \item \textit{There is a lack of a benchmarking tool that enables businesses to identify which countries are becoming blockchain ``friendly'' to host their operations, and establish their investment decisions.}  
\end{itemize}

As a response to the aforementioned observations, BRI \cite{vlachos2019algorithmic} provides a numerical metric (referred to as the blockchain \textit{readiness score}) in measuring the maturity/readiness levels of a country in adopting blockchain and cryptocurrencies.
 

Blockchain technology holds the potential to enable global societal shifts with disruptive effects both on nations and the global economy \cite{shin2020socio}. These shifts are likely to have a greater impact when converge with other disruptive technologies from Industry 4.0 \cite{bodkhe2020blockchain}. However, this transformational change is not likely to occur automatically but it will require a proactive collaboration among stakeholders from the ecosystem, governments, policy-makers, and regulators. The work presented in this paper is motivated by the different dynamics and approaches countries have taken or plan to execute towards the technology and adoption of cryptocurrencies. In their attempts to adapt to change, countries are trending towards proposing an increased political and regulatory control over blockchain technology and the cryptocurrency cyberspace. 

At the same time, many companies, investors, and start-ups from various industries have been experimenting with the technology and have identified unique business opportunities emerged from the ecosystem. However, a challenge remains in identifying the most ``enabling environment'' to host their blockchain-based activities and business propositions implemented with blockchain technology. Although nations are faced with potential new opportunities emerging from embracing disruptive technologies such as blockchain technology, at the same time they are exposed to new types of risks, that need to be identified and managed. Risk management in such an ecosystem is likely to consider adaptations to existing governance frameworks, policy protocols (\eg taxation, fraud protection), investment and financial models. At the same time, risk management proposals should aim towards promoting a balance between regulation and innovation that incentivizes further development of the technology, and societal engagement.

\subsection{Summary of Contributions}
\label{ss:contributions}

This paper extends our previous contribution \cite{vlachos2019algorithmic} of an algorithmically derived Blockchain Readiness Index (BRI) by accurately estimating the readiness score per country even in the presence of several missing indicators. Section \ref{s:model} discusses the details of our enchanced model for deriving the BRI that makes no assumptions on no missing values for indicators. More specifically, this paper contributes the following: (a) a robust indexing model for the proposed BRI, that provides accurate estimates on missing values for indicators; and (b) an empirical evaluation for measuring the effectiveness of our numerical estimates (on linear/sigmoid weights and weighted similarities) adopting a classification framework.

The first version of our BRI \cite{vlachos2019algorithmic} assumed the existence of numerical values to all indicators characterising each country. Thus, the methodology was producing limited rankings in cases where numerical values were missing from several indicators. In this work, we present a set of weighting schemes (as described in Section \ref{s:model}) to estimate the values from missing indicators, based on the assumption that countries that share similar blockchain-related indicators are likely to exhibit similar blockchain maturity/readiness. Thus, enabling our  BRI to estimate indexing scores per country even when information is partially missing.

Overall, our research work aims to provide business intelligence and insights on the opportunities and risks -- identified by the fast-moving blockchain environment -- to business executives, individuals, and start-ups. The proposed BRI takes into consideration several indicators in assessing the enabling conditions for supporting such actors in their decision making in establishing blockchain related businesses. More specifically, in identifying which country holds the most enabling conditions to host their blockchain-based projects and activities. The objectivity aspect of the BRI, which is influenced by the numerical scores feeding the system, remains outside the scope of this paper. This work focuses on the technical implementation details of algorithmically deriving such an index based on several indicators. We note that our implementation is generic and can work with any number of countries and indicators on the condition that signals from the indicators are provided as numerical scores.


\section{Related Work}
\label{s:back}

To date, there is limited research done in constructing an algorithmic readiness index to assess the level of preparedness for blockchain technology and cryptocurrencies among nations. This section briefly reports on well-known readiness indexes proposed in the literature for various industries and positions this work in the landscape of algorithmic readiness indexes for the disruptive blockchain technology. 

\subsection{Readiness Indexes}
\label{ss:back:readiness}

%
%
Readiness indexes are developed to provide a numerical representation of how engaged an examined item is -- such as a nation -- towards a specific subject matter. A number of technological indexes have been developed to date as an assessment of various technological metrics. The Networked Readiness Index (NRI) proposed by the Global Information Technology Report \cite{baller2016}, attempts to measure the propensity of nations to exploit the opportunities offered by Information and communications technology (ICT) developments. The latest release of the NRI \cite{dutta2019NRI} reports 121 nations based on four pillars: Technology, People, Governance and Impact. The general aim of NRI is to mesure how ICT is penetrating into countries and what is the impact
on their economies and on the ability to fulfill a set of defined Sustainable Development Goals (SDGs). Our research aims on developing a similar index for measuring the level of readiness of countries with regards to blockchain technology and cryptocurrencies. The proposed BRI is set to provide a general perspective on the current patterns and approaches nations taking with regards to blockchain and cryptocurrencies. In doing so, the BRI is  focusing among others on various features, such as, regulation regimes, research competences and awareness of the local blockchain community. 
Furthermore, indexes aiming to provide rankings of nations/regions regarding their readiness to become front-runners on new technological advancements. The best known examples include the Autonomous Vehicles Readiness Index (AVRI) \cite{threlfall2018autonomous} which provides a tool for assessing the level of readiness for autonomous vehicles; the Automation Readiness Index (ARI) \cite{eco2020} which assesses the level of preparedness for intelligent automation; and the Smart Industry Readiness Index (SIRI) published by the Singapore Economic Development Board \cite{siri2020} which aims to support manufacturers to assess and compare their maturity levels against Industry 4.0 and global industry benchmarks. Other indexes that attempt to report the readiness and maturity levels for Industry 4.0 are discussed in \cite{basl2018analysis}. 

Although the landscape of indexes covers a wide spectrum from various industries, a significant research gap exists as far as developing blockchain indexes and maturity metrics. The aim of this study is to construct an index that algorithmically aggregates data from various indicators per country into a single numerical score. This numerical score characterises the level of readiness a country is in terms of blockchain technology and cryptocurrencies.

The world is now scratching the surface of Blockchain technology \cite{zheng2017overview}. Despite the fact that blockchain is expected to impact the online world by enabling decentralized applications, smart contract transaction and be combined with other emerging technologies (\eg Internet Of Things and Artificial Intelligence), handling this technology is currently challenging for nations \cite{batubara2018challenges,shin2020socio}. Legislation and technological innovation are developed at a variable frequency within each nation. In addition, the rate of local blockchain awareness or engagement, and research capabilities differs per country. It is therefore, challenging for businesses, investors, and startups to examine the landscape of nations and identify the most ``enabling environment'' to host their blockchain-based activities.

A relevant study introduced the Blockchain and Cryptocurrencies Regulation Index (BCRI) \cite{flying2018}. In this work, authors describe a methodology for assessing the degree of enabling indicators and controls of cryptocurrencies for various countries. The ranking presented provides a single numerical score for each country, as well as, a general classification. The rationale of their approach is based on annotating each country with a positive or negative assessment on different enabling environments in the following dimensions: (i) legal environment, (ii) political environment, and (iii) infrastructure environment. However, the implementation details of the framework are abstracted without a clear indication on whether the index is algorithmically and/or dynamically derived or how the framework is treating missing values from indicators. 

In contrast, our methodology is dynamic and not restricted to any number of indicators. As long as these indicators are expressed as numerals the index can be derived. In addition, we have extended the first version of our BRI to effectively deal with countries that have indicators with missing values. The general formula used to calculate the values for the BCRI index is based on a weighted average. To compute the final score for our index, our methodology characterises each country by a feature vector consisting of indicators. It then makes use of \textit{cosine similarity} to derive the similarity of each country with the ``reference country'' \cite{vlachos2019algorithmic}. Finally, with this work we are contributing a classification framework for conducting an empirical evaluation of the BRI index, something which has not been discuss in the BCRI index.   


\section{Indicators of Blockchain Readiness}
\label{s:indicators}

\begin{figure}[htbp]
\centerline{\includegraphics[width=1\textwidth]{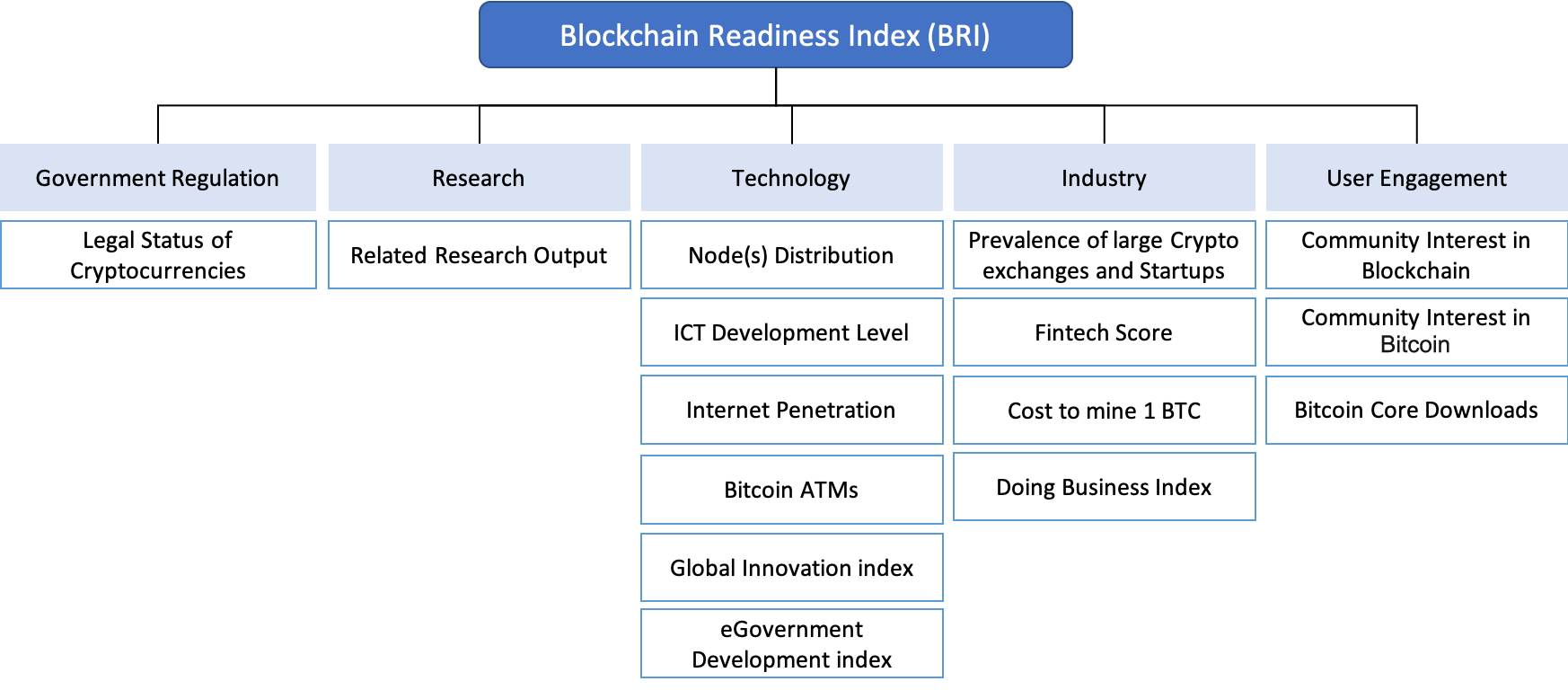}}
\caption{Indicators considered by the proposed Blockchain Readiness Index.}
\label{fig:indicators}
\end{figure}

This section provides a brief description of the indicators considered by BRI inherited from our previous work \cite{vlachos2019algorithmic}. Similarly to our previous work such indicators are categorized into groups referred to as \textit{enablers}. Figure \ref{fig:indicators} provides an overview of the indicators considered by the proposed updated version of the BRI. 

To define the indicators used for constructing the BRI we have followed standard procedures suggested by \cite{rust1994reliability}. According to Rust \etal rankings shall be based upon relevant and reliable indicators indicating a high level of data availability and theory for each \cite{rust1994reliability}. However, it is  challenging to gather data sources which would provide data for all indicators for all countries. Thus the first version of the BRI allowed us to report only those countries in which we had data for all indicators. This was a limitation since some countries with partial data could not be considered as part of the index. Moreover, we had to limit the feature set used to describe each country to the lowest common denominator of indicators. This paper deals with this limitation (see Section \ref{s:model}) by proposing a robust model for estimating such missing indicators. The proposed model is based on the assumption that countries which share similar blockchain-related indicators tend to exhibit similar blockchain readiness.

\subsection{Technology Indicators}
\label{ss:indicators_tech}
The significance of technological innovation within a governmental ecosystem is emphasized by \cite{dolfsma2013government}. For this enabler the following indicators are considered.

\vspace{0.2cm}

{\bf Node Distribution:}
The estimation of the size of Bitcoin (Global Bitcoin nodes distribution\footnote{\url{https://bitnodes.io/}}) and Ethereum (The Ethereum Network and Node Explorer\footnote{\url{https://www.ethernodes.org/}}) is derived by identifying all the reachable nodes within countries. For this version of the BRI we consider only Bitcoin and Ethereum as these public decentralized networks are widely transacted since their Genesis block. We note that our proposed methodology can be expanded to consider other data sources for all blockchain protocols available.

{\bf ICT Development Level:}
The ICT Development Index (IDI) is an index published by the United Nations International Telecommunication Union based on internationally agreed ICT indicators. In brief this composite index combines several indicators into one benchmark for monitoring the development of ICT over countries. The intuition is that this index can lead to signals indicating room for innovation towards Blockchain-specific activities, especially for high ranked countries.

{\bf Internet Penetration:}
Internet penetration rates indicate the prospect of Blockchain adoption within nations. Internet World Stats\footnote{\url{https://www.internetworldstats.com/}} is a useful source for country and regional statistics, international online market research, the latest Internet information, world Internet penetration data, world population statistics, telecommunications information reports, and Facebook statistics by country. This information may not be directly related to blockchain engagement but high internet penetration rates indicate a positive sign.

{\bf Bitcoin ATMs launched:}
Bitcoin ATMs is a convenient first introduction for many people to the Bitcoin ecosystem. Installation rates indicate the rate at which a country is embracing Bitcoin and how easy it is for the population to deposit and withdraw Bitcoin in exchange for cash. Installation rates are obtained by Coin ATM Radar\footnote{\url{https://coinatmradar.com/}}.

{\bf Global Innovation Index:}
The Global Innovation Index (2018) provides comprehensive metrics regarding the innovation performance of several countries worldwide. 80 indicators have been developed and evaluate the political environment, education, infrastructure and business sophistication.

{\bf eGovernment Development Index (EGDI):}
The EGDI \cite{UN2018} assesses eGovernment development at a national level and is composed by three components: online service index, telecommunication infrastructure index and the human capital index. This index provides signals on the transformation towards sustainable and resilient societies for each country.


\subsection{Industry Indicators}
\label{ss:indicators_industry}
The penetration of blockchain technology to the industry and emerging businesses is significant. Industry engagement and willingness to develop the blockchain sector of the economy, can play a dynamic role towards a society’s social and economic status in the future \cite{friedlmaier2018disrupting}.

\vspace{0.2cm}

{\bf Prevalence of top 100 Cryptocurrency Exchanges by Volume:}
The number and volume of cryptocurrency transactions is able to indicate the degree in which an economy is financed from the blockchain industry. The top 100 cryptocurrency exchanges by volume can be derived by the metrics of Coinmarketcap\footnote{\url{https://coinmarketcap.com/}}.

{\bf Fintech Score:}
The Findexable Global Fintech Index City Rankings report (2020)\footnote{\url{https://findexable.com/}} presents the results of an index algorithm which ranks the fintech ecosystems of more than 230 cities, 65 countries and 7000 fintech companies. 

{\bf Cost to mine 1 Bitcoin:}
This metric identifies how likely a country is to host mining operations. Energy consumption and mining facilities are developed on top countries. This provides a signal on how engaged the ecosystem is within a local community.

{\bf Doing Business Index:}
Doing Business Index (2020)\footnote{\url{https://www.doingbusiness.org/en/rankings}} is taking into consideration the ease of starting a business, dealing with construction permits, getting electricity, registering property, getting credit, protecting minority investors, paying taxes, trading across borders, enforcing contracts, and resolving insolvency within a country.


\subsection{User Engagement Indicators}
\label{ss:indicators_user}
Individuals and local startups have initiated various community efforts. This enabler attempts to capture positive or negative signals on how community is reacting to blockchain technology and cryptocurrencies. 

\vspace{0.2cm}

{\bf Community interest in Blockchain:}
This indicators captures the increase over time of the number of Web search keywords that include the term ``Blockchain''. This indicates a trend within a country’s interest to the technology.

{\bf Community interest in Bitcoin:}
Similarly to the above, the increase over time of the number of Web searches that include the term “Bitcoin”. This indicates a trend within a country’s interest to Bitcoin as a concept and as a cryptocurrency. Other keywords such as ``distributed ledgers'' or ``decentralization'' are also considered.

{\bf Bitcoin Core downloads:}
The total number of Bitcoin Core downloads indicates local engagement and interest and a rough approximation of where most Bitcoin users are located. The data period range can be adjusted; for the purposes for this study we use data for the previous 365 days.


\subsection{Government Regulation Indicators}
\label{ss:indicators_gov}

Governments aiming to boost innovation through breakthrough technologies such as blockchain with many attempting to experiment with more decentralized and transparent form of governance operations \cite{olnes2016beyond}. The regulatory indicators are used to capture the current landscape with regards to the various regulatory circumstances that are likely to effect a country’s perception to becoming a blockchain hub.

{\bf Cryptocurrency Regulation Analysis:}
Data reported by the Worldwide Cryptocurrency Regulation Analysis (2020)\footnote{\url{https://cointobuy.io/}}, are taken into consideration along with the following metrics:

\begin{enumerate}[label=\roman*.]
  \item Legality of Bitcoin,
  \item ICOs restrictions,
  \item ICOs registration locations,
  \item Exchanges locations, and
  \item User voting (public opinion).
\end{enumerate}


\section{Robust Indexing: Proposed Model}
\label{s:model}
This section discusses the proposed model leveraged by our algorithmic BRI in deriving the index even in cases of countries are missing values for some indicators. The indicators are used to construct a feature vector that is used to characterise each country. For our work this feature vector is an $N$-dimensional vector for each country (denoted by $c$) consisting of numerical values for each indicator, $k$.


Let us assume a dataset consisting of countries for which their respective BRI scores should be estimated. On the basis of those scores, the countries can be ranked (typically in descending order) resulting into a readiness index. At the abstract level, this process is composed by the following steps.

In the first step, any missing BRI indicators should be estimated for each country.
During the second step, the BRI indicators (either initially available or estimated), are exploited for computing a single BRI score for each country. This score is meant to quantify the blockchain readiness of the corresponding country. The models proposed for executing the aforementioned steps are presented in Section~\ref{ss:model-mis}, and Section~\ref{ss:model-rank} respectively.

\subsection{Estimation of Missing Indicators}
\label{ss:model-mis}

Consider a country $c$ for which the values of one or more indicators are missing.
Let us denote with $\hat{I}_{c,k}$ the value of the $k$--th indicator of $c$ which is estimated as:

\begin{equation}
\label{eq:ind-estimated}
\hat{I}_{c,k} =
\frac{1}{\mid\!T_{c}\!\mid}\sum_{t \in T_{c}} {I}_{t,k},
\end{equation}

where, $T_{c}$ stands for the set of $c$'s most similar countries computed according to (\ref{eq:cosine_sim}), and ${I}_{t,k}$ is the value of the $k$--th indicator of country $t$ being member of $T_{c}$.

In general, similarity computation is used as a tool for handling the cases of missing operators providing robustness to the proposed approach.
The effect of $\mid\!T_{c}\!\mid$ in terms of performance is reported in Section \ref{s:eval}.

The following example describes the overall procedure for a hypothetical scenario where the maximum number of indicators equals to three.
In this context, assume a country $c_0$ which is missing the third indicator,
while the respective vector of indicators is $[0.25, 0.30, 0.00]$.
Note that for the case of missing indicators, the respective non-available values are substituted by zeros values (representing the absence of information).
For estimating the value of the missing indicator of $c_0$,
the countries having all indicators are considered.
Assume three such countries denoted as $c_1$, $c_2$, and $c_3$.
Let their vectorized indicators be $[0.17, 0.20, 0.20]$,
$[0.15, 0.18, 0.35]$, and
$[0.28, 0.16, 0.30]$, respectively.
The next step is to compute the similarity between $c_0$ and those countries.
By applying the cosine similarity, the following similarity scores are yielded:
$0.795$, $0.556$, and $0.686$, respectively.
In order to estimate the value of $c_0$’s missing indicator,
a number of top similar countries (with respect to $c_0$) are considered.
For the purposes of the present example, this number is set to two.
Thus, $c_1$ and $c_3$ should be considered as they constitute the two most similar countries of $c_0$.
The values of the third indicator of $c_1$ and $c_3$ are taken into account through an averaging operation (i.e., $\frac{0.20+0.30}{2}=0.25$).
This results into the following vector of indicators for
$c_0$: $[0.25, 0.30, 0.25]$ where the zero value was substituted by $0.25$.

\subsection{Ranking}
\label{ss:model-rank}

Consider the ideal country $\widetilde{c}$ and some country $c$. For the purposes of BRI the feature vector of the ideal country exhibits the best possible value for each indicator \cite{vlachos2019algorithmic}.

For example, assume two countries characterized by three indicators.
Let the vectorized representation (i.e., feature vector) of those countries be as follows: $[0.52, 0.63, 0.19]$ and $[0.71, 0.25, 0.80]$.
The feature vector of the ideal country is computed by applying the maximum operator element-wise resulting into $[0.71, 0.63, 0.80]$.

The ranking of $c$ is conducted with respect to a score $S_{c}$ which is computed as:
\begin{equation}
\label{eq:rank}
S_{c} = f(\widetilde{c},c)g_{c},
\end{equation}
where, $f(\widetilde{c},c)$ denotes the cosine similarity between ideal country $\widetilde{c}$ and some other country $c$.
In general, the cosine similarity is widely-used measurement
that has been utilized in various areas including
semantic web (e.g., \cite{christodoulou2015structure})
and natural language processing (e.g., \cite{iosif2015similarity}),
especially for tasks related to unsupervised machine learning. 

The similarity between $\widetilde{c}$ and $c$, $f(\widetilde{c},c)$,
is estimated as the cosine of their respective vectorized indicators' values
\footnote{
In general, any similarity (or distance) metric can be used.
In this work, we have also experimented with Euclidean distance without observing any improvement for the experiments reported in Section~\ref{s:exps}.
}
:
\begin{equation}
\label{eq:cosine_sim}
    f(\widetilde{c},c)=
    \frac{\sum_{i=1}^{N}I_{\tilde{c},i} I_{c,i}}
    {\sqrt{\sum_{i=1}^{N}(I_{\tilde{c},i})^2}\sqrt{\sum_{i=1}^{N}(I_{c,i})^2}},
\end{equation}

where, $I_{\tilde{c},i}$ and $I_{c,i}$ are the values of the $i$--th indicator of
$\widetilde{c}$ and $c$, respectively, while $N$ stands for the number of indicators.

The similarity scores computed by (\ref{eq:cosine_sim}) lie in $[0, 1]$,
with $0$ and $1$ denoting zero and absolute similarity, respectively. 

The $g_{c}$ constituent of (\ref{eq:rank}) is defined as:

\begin{equation}
\label{eq:ind-coverage}
g_{c} = \frac{N-n_{c}}{N},
\end{equation}

where, $N$ is the number of indicators and $n_{c}$ denotes the number of $c$'s
null indicators.

Both $f(\widetilde{c},c)$ and $g_{c}$ range in the $[0, 1]$ interval.

An alternative scheme of $S_{c}$, denoted as $S_{c}^{'}$, is defined as follows
($g_{c}$ is substituted by a sigmoid function in which $g_{c}$ also appears as a parameter): 

\begin{equation}
\label{eq:rank-alt}
S_{c}^{'} = f(\widetilde{c},c)
\left(1+\left(\frac{g_{c}(1-\gamma)}{\gamma(1-g_{c})}\right)^{-2}\right)^{-1},
\end{equation}

where, $g_{c}$ is computed according to (\ref{eq:ind-coverage}), as in the case of (\ref{eq:rank}).

As in the initial scheme, $S_{c}^{'}$ computes scores that within $[0, 1]$.
$\gamma$ is constant that typically takes values as $0 \le \gamma \le 1$
and it can be used for centering the sigmoid function in the the $[0, 1]$ domain. The basic idea that underlies (\ref{eq:rank-alt}) is the weighting of $f(\widetilde{c},c)$ according to a non-linear scheme. In Section \ref{s:eval}, the performance for various values of $\gamma$ is presented and discussed.

Overall, either $S_{c}$ score or $S_{c}^{'}$ score can be used for computing the final BRI \ie the ranking of countries according to their blockchain readiness.
A comparison between $S_{c}$ and $S_{c}^{'}$ is also reported in Section \ref{s:eval}.


\section{Experimental Data and Setup}
\label{s:exps}
This section presents the experimental data used along with the setup of the experiments. In addition, the evaluation process and metric are described
followed by two experimental baselines.

The dataset used in this work is summarized in Table \ref{tab:dat}.

\begin{table} [!ht]
\centering
\begin{tabular}{|c|c|}
\hline
Number of countries annotated as ``high BRI'' & 45 \\
\hline
Number of countries annotated as ``mid BRI'' & 55 \\
\hline
Number of countries annotated as ``low BRI'' & 90 \\
\hline
\hline
Total number of countries & 190 \\
\hline
\end{tabular}
\caption{Overview of experimental dataset.} 
\label{tab:dat}
\end{table}

In total, there are $190$ countries categorized with respect to the following BRI levels (labels): (i) ``high BRI', (ii) ``mid BRI'', and (iii) ``low BRI''. Each country was assigned (in the form of annotations) a category grounded on the rational decisions made by human experts
\footnote{
The annotations of three experts were used and an overall annotation was compiled considering the three individual annotations.
The overall annotation (i.e., after consolidating the three individual annotations)
was used for the experimental part of this work.
}
.

More specifically, $45$ out of $190$ were annotated as countries featuring ``high BRI'', while $55$ and $90$ countries were assigned the ``mid BRI'' and ``low BRI'' annotations, respectively.

\begin{figure*}[!htbp]
\begin{multicols}{2}
    \includegraphics[width=\linewidth]{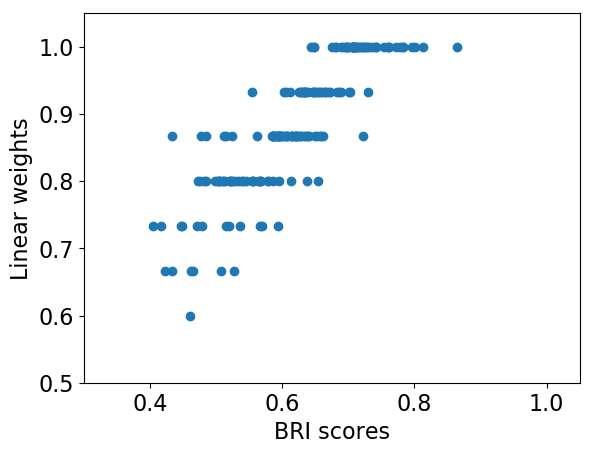}\par 
    \caption*{(a)}
    \includegraphics[width=\linewidth]{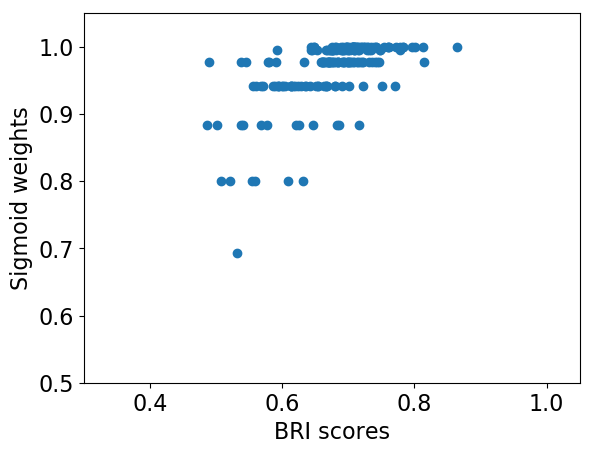}\par 
    \caption*{(b)}        
    \end{multicols}
\captionof{figure}{BRI scores for (a) linear and (b) sigmoid weighting schemes.}
\label{fig:scatter}
\end{figure*}

Figure \ref{fig:scatter} depicts the relationship between BRI scores and the respective weights according to the linear (see (a)) and sigmoid scheme (see (b)).

By definition, both scores and weights lie within the $[0,1]$ interval. We observed that for both schemes, linear and sigmoid, there is linear correlation between weights and BRI scores.
On another note, it is evident that a significant portion of the considered countries exhibit partial indicators (those countries are assigned weights being less than $1$).
The key difference between linear and sigmoid weighting schemes is that the former
(see Figure \ref{fig:scatter}(a)) yields a more distinctive separation of countries that are characterized by mid (around $0.50$) to high ($>0.85$) BRI scores.

The key parameters of the proposed models, defined in Section \ref{s:model},
are as follows:

\begin{enumerate}
    \item Weighting scheme used for BRI: linear or sigmoid (use of (\ref{eq:rank}) or (\ref{eq:rank-alt}))
    \item $\gamma$ used in sigmoid weighting (see (\ref{eq:rank-alt}))
    \item Given a country $c$, the number of $c$'s most similar countries (see $\mid\!T_{c}\!\mid$ in (\ref{eq:ind-estimated}))
\end{enumerate}

The performance for various values set to the above parameters are reported in Section \ref{s:eval}.
Next, we describe an experimental setup that is formulated in the context of supervised learning.

\begin{itemize}
    \item \textbf{Features:} Two features were utilized, namely, the weighting scores (linear or sigmoid) and the respective BRI scores.
    \item \textbf{Classification task:} The experimental task of this work was defined as a classification problem. Given the aforementioned features for a country, the task is to assign a label to it \ie``high BRI'' or ``mid BRI'' or ``low BRI''.
    \item \textbf{Classification models:} We have experimented with various classifiers including Naive Bayes (NB), Support Vector Machines (SVM), and Random Forest (RF).
    \item \textbf{Evaluation process:} A $10$-fold cross validation process was applied using the dataset presented in Table \ref{tab:dat}.
    \item \textbf{Evaluation metric:} Classification accuracy was used for evaluating the performance of classifiers. It is computed as the percentage of the correctly classified test instances. Specifically, in the framework of $10$-fold cross validation, the performance is reported in terms of average classification accuracy by averaging the classification accuracy scores that correspond to each fold.
 \end{itemize}
 
All experimental results are reported only for the case of SVM. This is because very similar performance was achieved for the case of NB, while the use of RF resulted in lower classification accuracy compared to SVM and RF.
Regarding SVM, we used the SMO algorithm \cite{platt1998fast} with the following configuration:
(i) use of polynomial kernel (degree of polynomial: $1.0$),
(ii) the complexity parameter: $1.0$,
(iii) epsilon for round-off error: $1 \times 10^{-12}$,
(iv) tolerance parameter: $0.001$.
The aforementioned categorization from human experts was used as ground truth for training/testing purposes.

Within the present classification-based experimental framework, the following classification baselines were adopted:

\begin{itemize}
\item \textbf{Baseline $1$:} Assume no classification model. Given any (unknown) country, always assign to it the label of the most populous class
(\ie ``low BRI'', see Table \ref{tab:dat}).
\item \textbf{Baseline $2$:} Use no weighting \ie $g_{c}=1$ in (\ref{eq:rank}).
\end{itemize}

Baseline $1$ was meant for testing the usefulness of a classification model that goes beyond the naive  most--populous--class strategy. Baseline $2$ was adopted for checking whether $f(\widetilde{c},c)$ needs weighting. 

\section{Evaluation Results} 
\label{s:eval}
This section, reports the evaluation results derived in terms of average classification accuracy with respect to the granularity schemes presented in Table \ref{tab:granularity}.
\begin{table} [!ht]
\begin{center}
\begin{tabular}{|c|c|}
\hline
Granularity scheme & Class labels \\
\hline
\hline
$3$-class & ``high BRI'', ``mid BRI'', ``low BRI'' \\
\hline
$2$-class & ``high BRI'', ``low BRI'' \\
\hline
\end{tabular}
\end{center}
\caption{Classification of granularity schemes.}
\label{tab:granularity}
\end{table}

The $2$-class granularity scheme focuses on the ``high--vs.--low'' discrimination
based on the hypothesis that there are cases where the discrimination of two
ends of the BRI spectrum is adequate (as opposed to the case of trying to build a more fine-grained model). The results reported were obtained using the two features: (i) weighting scores (linear or sigmoid), and (ii) the respective BRI scores.

Table~\ref{tab:results}, summarizes the performance for both linear and sigmoid schemes. The performance is reported for the $3$-class and $2$-class granularity schemes. In addition, the performance of the two baselines is included for comparison purposes.

\begin{table} [!ht]
\begin{center}
\begin{tabular}{|c||c|c|}
\hline
Features & Linear scheme & Sigmoid scheme \\
& (use of $S_{c}$) & (use of $S_{c}^{'}$) \\
\hline
\hline
\multicolumn{3}{|c|}{3-class}\\
\hline
Baseline 1 & $47.4$ & $47.4$ \\
\hline
Baseline 2 & $48.4$ & $48.4$ \\
\hline
Proposed features & ${\bf 66.8}$ & $62.6$ \\
\hline
\hline
\multicolumn{3}{|c|}{2-class}\\
\hline
Baseline 1 & $66.7$ & $66.7$ \\
\hline
Baseline 2 & $68.9$ & $68.9$ \\
\hline
Proposed features & ${\bf 89.6}$ & ${\bf 89.6}$ \\
\hline
\end{tabular}
\end{center}
\caption{Classification accuracy (\%) for 3-class and 2-class task
(for $\mid\!T_{c}\!\mid$=10, $\gamma=0.7$).}
\label{tab:results}
\end{table}

Firstly, it is observed that the use of the proposed features outperforms
the two baseline approaches. This holds for both weighting and granularity schemes.
Regarding the $3$-class granularity scheme, the highest classification accuracy ($66.8\%$) is yielded by the linear scheme. The sigmoid scheme obtains slightly lower accuracy ($62.6\%$). For the $2$-class granularity scheme, the two weighting schemes achieve identical performance being equal to $89.6\%$

\begin{table} [!ht]
\begin{center}
\begin{tabular}{|c||c|c|c|c|c|c|c|c|c|}
\hline
$\gamma$ & 0.1 & 0.2 & 0.3 & 0.4 & 0.5 & 0.6 & 0.7 & 0.8 & 0.9\\
\hline
\hline
3-class & 50.0 & 51.1 & 52.1 & 53.2 & 53.9 & 53.7 & 62.6 & 61.6 & {\bf 64.2}\\
\hline
2-class & 83.0 & 84.4 & 83.7 & 85.9 & 87.4 & 88.9 & {\bf 89.6} & 88.9 & 88.9\\
\hline
\end{tabular}
\end{center}
\caption{Classification accuracy (\%) for various $\gamma$ values of the sigmoid scheme (for $\mid\!T_{c}\!\mid$=10).
}
\label{tab:results-sigmoid-param}
\end{table}

Table~\ref{tab:results-sigmoid-param}, presents the performance for various values of the $\gamma$ factor that is used in sigmoid weighting scheme (see (\ref{eq:rank})). As before, this is shown for both classification granularity schemes. In terms of performance scores, the $3$-class scheme exhibits greater variance when compared to the $2$-class scheme. Regarding the $3$-class scheme, the highest classification accuracy ($64.2\%$) is achieved when $\gamma=0.9$. For the $2$-class case top performance ($89.6\%$) is yielded for $\gamma=0.7$.

\begin{table} [!ht]
\begin{center}
\begin{tabular}{|c||c|c|c|c|c|c|c|c|c|}
\hline
$\mid\!T_{c}\!\mid$ &1 & 2 & 3 & 5 & 10 & 15 & 20 & 30 & 40 \\
\hline
\hline
3-class & 65.8 & 65.8 & 64.2 & 64.7 & {\bf 66.8} & 66.3 & 66.8 & 65.8 & 64.7\\
\hline
2-class & 91.1 & 91.1 & {\bf 91.9} & 91.1 & 89.6 & 91.1 & 91.1 & 90.4 & 90.4\\
\hline
\end{tabular}
\end{center}
\caption{Classification accuracy (\%) for various
$\mid\!T_{c}\!\mid$ values using the linear scheme.
}
\label{tab:results-linear-param}
\end{table}

The classification accuracy for various values of $\mid\!T_{c}\!\mid$, which appears in (\ref{eq:ind-estimated}) and denotes the number of $c$'s most similar countries, is shown in Table~\ref{tab:results-linear-param}.

The accuracy scores are presented with respect to the two classification granularity schemes. For the case of $3$-class, the highest performance ($66.8\%$)
is yielded by the use of $\mid\!T_{c}\!\mid=10$. Regarding the $2$-class granularity scheme, the top classification accuracy ($91.9\%$) is achieved for $\mid\!T_{c}\!\mid=10$.

A number of indicative classification outputs
are as follows:
Canada and France (``high BRI''),
Belarus and Greece (``mid BRI''), and
Andorra and Maldives (``low BRI'').
Those outputs were computed by the model
that yielded the top classification accuracy ($66.8$)
for the $3$-class case.

Furthermore, in Table \ref{tab:confusion-mat}
we present the confusion matrix that corresponds
to the top-performing setting ($91.9\%$ in Table \ref{tab:results-linear-param}).
This details the performance of the classifier in term of misclassification types
(``low BRI instead of high BRI'', and ``high BRI instead of low BRI'').
\begin{table} [!ht]
\begin{center}
\begin{tabular}{|c|c|}
\hline
Low BRI & High BRI\\
\hline
\hline
88 (correct) & 2 (error)\\
\hline
9 (error) & 36 (correct)\\
\hline
\end{tabular}
\end{center}
\caption{Confusion matrix of the top-performing classification setting:
number of correctly/erroneously classified countries.}
\label{tab:confusion-mat}
\end{table}
For the case of ``high BRI'', the classification outcome is correct for $36$ countries,
while only two countries were misclassified.
Regarding the ``low BRI'' case, $88$ countries were correctly classified
and nine countries were erroneously put under the opposite category.
Since the majority of predictions made by the classifier are correct,
we list the countries for which the classification outcome was not accurate
(values $2$ and $9$ in the confusion matrix).  
The countries that were erroneously classified as ``high BRI'' are as follows: 
(i) Albania, and (ii) Bangladesh. 
The countries erroneously classified as ``low BRI'' are: 
(i) Bahrain, (ii) Croatia, (iii) Hong Kong,
(iv) Liechtenstein, (v) Saudi Arabia, (vi) Serbia,
(vii) Turkey, (viii) Ukraine, and (ix) Venezuela. 
An indicative run of our algorithmic readiness index resulted to the following ranking\footnote{Ranking: \url{https://bit.ly/3neHbc0}}.

Overall, excellent classification accuracy (up to $91.9\%$) was achieved for the case of $2$-class granularity scheme; which focuses on the discrimination of
``high BRI'' vs. ``low BRI'' countries. The $3$-class granularity scheme poses a more difficult classification problem and, as anticipated, a lower performance score ($66.8\%$ classification accuracy) was obtained. This difficulty can be attributed on the presence of countries that lie in the middle of the BRI spectrum.
Regarding the hardest classification task, \ie the one based on the $3$-class granularity scheme, the linear weighting scheme appears to performs better than the sigmoid scheme. This observation can be intuitively explained by inspecting Figure \ref{fig:scatter}. Specifically, the distribution of features values for the linear case (Figure \ref{fig:scatter}(a)) makes the three classes of interest more separable compared to the sigmoid case (Figure \ref{fig:scatter}(b)).


\section{Conclusions and Future Work}
\label{s:concl}

In this work, we proposed an updated version of our previous approach on an algorithmic computation of a blockchain readiness score (referred to as Blockchain Readiness Index -- BRI) at the country level. BRI utilizes a set of indicators being related to the blockchain maturity exhibited by the countries under investigation. The core contribution of this work is a technique for estimating the BRI in the presence of missing indicators (extending our previous approach which requires no missing indicators). We coined the term ``robust BRI'' to refer to the ability of estimating BRI scores even when information for the indicators is missing. 

This improves our algorithmic approach for deriving the BRI, since a significant number of countries under consideration is characterized by partial indicator sets. With this additional optimization, our technique covers a larger (in terms of coverage of countries) list of countries and could be used to algorithmically derive extended indexes.

The estimation of missing indicators was based on the assumption that countries that share similar blockchain-related indicators are likely to exhibit similar blockchain readiness. Under this reasoning, we proposed two weighting schemes for refining the initial similarity estimates, that constitute the building block of BRI, namely, linear and sigmoid weighting. In order to evaluate the effectiveness of the core numerical estimates (linear/sigmoid weights and weighted similarities) a classification-based framework was adopted utilizing supervised learning where those estimates were used for training several classifiers.

It was experimentally shown that the proposed classification features
significantly outperform the baseline approaches for both classifications tasks:
(a) ``high BRI'' vs. ``mid BRI'' vs. ``low BRI'' (also referred to as $3$-class task), and
(b) ``high BRI'' vs. ``low BRI'' (also referred to as $2$-class task).
Especially for the latter task, up to $91.9\%$ classification accuracy was achieved.
This experimentally justifies the effectiveness of the proposed approach regarding the estimation of missing blockchain indicators. In addition, it was found that the weighting of the similarity scores plays a critical role. The use of no weighting (Baseline 2) resulted in poor performance.

Regarding the investigated weighting schemes, the linear one appeared to perform better than the sigmoid scheme for the $3$-class task. This observation is important from the perspective of learning the model since the linear weighting scheme does not impose parameter learning (in contrast to the sigmoid scheme which includes a number of parameters that should be estimated/tuned). For the $2$-class task, both schemes exhibited identical performance.
Regarding the sigmoid scheme, the observation that the top classification
accuracy scores were obtained for $\gamma=0.9$ and $\gamma=0.7$
(for $3$-class and $2$-class tasks, respectively)  
suggests that the centering of the sigmoid function within the $[0,1]$ domain,
which corresponds to $\gamma=0.5$, is not the optimal configuration.
Regarding the number of most similar countries utilized during the estimation of a missing indicator value, it was interestingly observed that the proposed approach robust when relatively few countries are taken into considerations. Specifically, the best performing configurations correspond to $10$ and $3$ countries for $3$-class and $2$-class tasks, respectively.
Another key finding is that both Naive Bayes and Support Vector Machines resulted with almost identical performances. This is an indication that the pattern nonlinearities which, in principle, are captured by the latter are not necessary for this particular task.

In addition, this task by its nature makes the application of (data-demanding) deep learning algorithms quite challenging. That is, the number of countries which constitute the training samples is quite limited compared to other datasets that are typically used in deep learning. This realization led us to experiment with the aforementioned traditional models which yielded excellent performance.

For future work, we plan to further enhance the proposed approach for robust BRI estimation by investigating the weighted fusion of the indicators within the similarity computation phase. In this context, we will also explore the possible benefits of dimensionality reduction (\eg by applying Singular Value Decomposition) applied over the indicators space. In addition, we are currently working on the development of an online Web service for exposing and interacting with the estimated BRI. This can support customized functionality, such as the selection of indicators to be considered during the BRI estimation.

\bibliographystyle{apalike}
\bibliography{main}

\end{document}